\documentclass[reprint,prl,superscriptaddress]{revtex4-1}
\usepackage{ae}
\usepackage{amsmath}
\usepackage{amssymb}
\usepackage{amsfonts}
\usepackage{natbib}
\usepackage{soul}

\newcommand{\lpa}{\left(}
\newcommand{\rpa}{\right)}

\usepackage[T1]{fontenc}
\usepackage{bm}
\usepackage{color}
\usepackage{graphicx}
\usepackage[breaklinks=true]{hyperref}
\hypersetup{
  unicode=false,          
  pdftoolbar=true,        
  pdfmenubar=true,        
  pdffitwindow=false,     
  pdfstartview={FitH},    
  pdftitle={Effect of Partially Screened Nuclei on Fast-Electron Dynamics},    
  pdfauthor={Hesslow,Embréus,Stahl,DuBois,Papp,Newton, and Fülöp},     
  pdfsubject={Subject},   
  pdfcreator={Creator},   
  pdfproducer={Producer}, 
  pdfkeywords={keyword1} {key2} {key3}, 
  pdfnewwindow=true,      
  colorlinks=true,        
  linkcolor=blue,         
  citecolor=blue,         
  filecolor=blue,         
  urlcolor=blue           
}

\begin{document}

\title{Effect of Partially Screened Nuclei on Fast-Electron Dynamics}
\author{L.~Hesslow}
\affiliation{Department of Physics, Chalmers University of Technology, 41296 Gothenburg, Sweden} 
\author{O.~Embr\'eus}
\affiliation{Department of Physics, Chalmers University of Technology, 41296 Gothenburg, Sweden} 
\author{A.~Stahl} 
\affiliation{Department of Physics, Chalmers University of Technology, 41296 Gothenburg, Sweden}
\author{T.~C.~DuBois}
\affiliation{Department of Physics, Chalmers University of Technology, 41296 Gothenburg, Sweden}
\author{G.~Papp}
\affiliation{ Max-Planck-Institute for Plasma Physics, D-85748 Garching, Germany}
\author{S.~L.~Newton}
\affiliation{Department of Physics, Chalmers University of Technology, 41296 Gothenburg, Sweden}
\author{T.~F\"ul\"op}
\affiliation{Department of Physics, Chalmers University of Technology, 41296 Gothenburg, Sweden}
 \date{\today}

\begin{abstract}
  We analyze the dynamics of fast electrons in plasmas containing
  partially ionized impurity atoms, where the screening effect of
  bound electrons must be included.  We derive analytical expressions
  for the deflection and slowing-down frequencies, and show that they
  are  increased significantly
  compared to the results obtained with complete screening, already at
  sub-relativistic electron energies. Furthermore, we show that
    the modifications to the deflection and slowing down frequencies are of equal importance in
    describing the runaway current evolution.  Our results greatly affect fast-electron dynamics and have
  important implications, e.g.\! for the efficacy of mitigation
  strategies for runaway electrons in tokamak devices, and energy loss
  during relativistic breakdown in atmospheric discharges.
 \end{abstract}

\maketitle

\renewcommand{\thesection}{\Roman{section}}
\renewcommand{\thesubsection}{\Alph{subsection}} {\em Introduction.---}Fast electrons, having speeds well above the thermal speed of the
bulk plasma population, are ubiquitous in space and laboratory
plasmas.
An important process leading to such high-energy electrons  is the runaway mechanism.
Runaway electrons can be produced in the presence of an
accelerating electric field if it exceeds the critical value $E_c\!=\!n_e e^3 \ln\Lambda_0/4 \pi \epsilon_0^2 m_e c^2$~\cite{dreicer1959,connor}.
Situations where runaway electrons are believed to be important
include solar flares \cite{Holman1985}, atmospheric discharges \cite{dwyer,lehtinen},
laser-produced plasmas \cite{Parks2007}, as well as tokamak
disruptions~\cite{helander2002}.
In the latter, it is important to understand the dynamics of runaway electrons as they have the potential to seriously damage the
tokamak~\cite{Reux2015}.
This will be especially problematic in larger, future tokamak experiments, such as ITER~\cite{ITER}.
There, runaway-electron currents in excess of a megaampere are expected
to form if a disruption is not mitigated,  and the potential damage associated with such currents is larger than
in any present experiment~\cite{Hollmann2015}.
Therefore, reliable methods to deal with such currents are required~\cite{Boozer2015,Lehnen2015}.

One method to mitigate the detrimental effects of runaway-electron
beams is to dissipate them by injecting heavy
ions (impurities) such as argon or neon.
The impurity atoms become weakly ionized in the cold (few eV) post-disruption
plasma, and act to collisionally scatter particles in the high-energy
electron beam.
Experiments have shown that such injection of impurities with high atomic
mass can shift the energy distribution of the fast electrons towards
lower energies, 
to a much larger extent than  predicted by standard collisional theory~\cite{Hollmann2013,Hollmann2015}.
The discrepancy between measured and predicted dissipation increases with atomic mass,
even though the ions are usually only singly ionized in the cold,
post-disruption phase.
This is an indication that a fast electron is not simply deflected by
the Coulomb interaction with the net charge of the ion, but also probes its internal electron structure, so that the nuclear charge is not completely screened. The fast electrons can therefore be expected to experience higher collision rates against impurities, leading to a more efficient damping, in agreement with experimental observations.

To quantify the importance of 
this partial penetration of the electron cloud of the ion, 
 we compare the low-momentum-transfer limit of {\em complete screening} (CS) of the nuclear charge (i.e.~the electron interacts only with the net ion charge) to the high-energy limit of {\em  no screening} (the electron experiences the full nuclear charge), for both elastic and inelastic collisions.
 For elastic collisions, where the ion can be modeled as one entity~\cite{landauQM}, screening affects the interaction strength, which
is proportional to the charge squared. Compared to the limit of complete
screening, the case of no screening thus enhances the interaction
strength by a factor $X^2= (Z/Z_0)^2$, where $Z_0$ is the
ionization state and $Z$ is the charge number of the nucleus.  Furthermore, inelastic collisions (leading to excitation of the ion) can be treated as electron-electron interactions~\cite{bethe}, and thus increase the effective
electron density of the plasma (as experienced by the fast electron). The rate of electron-electron
collisions will therefore be of order $X$ larger, 
 as compared to the case where only free electrons are included.
 Since the factor  $X$ is large when weakly ionized high-atomic-number 
ions are present in the plasma, the effect of 
reduced screening 
on the collisional dynamics 
of fast particles 
can be significant for both elastic and
inelastic collisions.

To model the reduced-screening effect, a quantum-mechanical model must be
adopted. Because of the high speed of the incoming electrons, the
elastic collisions can be treated using the Born approximation
\cite{Kirillov, zhogolev}, which requires knowledge of the
electronic charge density of the impurity ion. The effect of inelastic
collisions can be modeled using Bethe's theory for the collisional
stopping power~\cite{mosher1975}.
A classical description of elastic collisions, combined with a
stopping-power formula for inelastic collisions~\cite{mosher1975}, was used in 
a test-particle approach in Ref.~\cite{martinsolis}. The results
indicated that the effect of the partially ionized impurities on the
runaway growth rate can be substantial.
However, without a quantum-mechanical treatment of the elastic
collisions and a solution of the kinetic equation for the fast-electron
distribution, the effect of screening on fast-electron dynamics
cannot be accurately quantified.
An extension to a quantum-mechanical treatment of elastic collisions
in the Born approximation was employed in Refs.~\cite{zhogolev,
  Kirillov, dwyer,lehtinen} using the Thomas--Fermi theory for the electron charge density,
 which is limited to intermediate distances from the nucleus,
and does not capture the shell structure of the ion~\cite{landauQM}. 
Based on the results of Ref.~\cite{Kirillov}, 
it was pointed
out in Ref.~\cite{Igitkhanov} that
the runaway generation rate 
is expected to be reduced.
However, an accessible analytical model of screening, that can be
incorporated in kinetic simulations of runaway-electron dynamics,
has not been derived. This is essential in order to combine all of the
important effects governing runaway dynamics into a tractable kinetic
model, which would enable a quantification of the effect of reduced screening on the
runaway-electron distribution.  

In this Letter we present a generalized collision operator which
accounts for the screening effect of bound electrons in collisions
between fast electrons and partially ionized impurities. We model
elastic electron-ion collisions quantum-mechanically in the Born
approximation, using density functional theory (DFT) to obtain the
electron-density distribution of the impurity ions.  This allows us to
determine the deflection frequency from first principles, without the
assumption of infinite nuclear charge used in the Thomas-Fermi model.
Furthermore, we employ stopping-power theory to describe inelastic
collisions with the bound electrons, and derive an expression for the slowing-down frequency. 
We demonstrate the effect of screening on the electron distribution
function via kinetic simulations.

{\em Collision operator.---}Small-angle collisions between species $a$ and $b$ can be
described by the Fokker-Planck operator
\cite{rosenbluth,akama}:
$  C^{ab}\!=\! -\nabla_k \left(f_a
  {\langle \Delta p^k \rangle_{ab} } \right) + \frac{1}{2} \nabla_k
  \nabla_l \left(f_a {\langle \Delta p^k \Delta p^l \rangle_{ab}}
  \right)$, 
  where $f_a$ is the distribution function of particle species $a$,  $\mathbf{p}\!=\!\gamma \mathbf{v}/c$ is the normalized momentum (with $\gamma$ the Lorentz factor), and $\Delta p^k$ the change in the $k$th component of the particle momentum in a collision. 
  The momentum averages  are given by 
${\langle \Delta
  p^k \!\cdots\! \Delta p^l \rangle_{ab} } = 
  \!\int\! d\mathbf{p}'
f_b(\mathbf{p'}) \int (d\sigma_{ab}/d\Omega) u \Delta p^k
\!\cdots\! \Delta p^l  d\Omega,
$
 where  
$u$ is the relative velocity between the particles,  and $d\sigma_{ab}/d\Omega$ is the differential cross-section.

When species $b$ has a Maxwellian distribution, the collision operator
can be simplified to
\begin{equation}
    C^{ab} = {\nu_D^{ab}} {\cal L}(f_a) +
    \frac{1}{p^2}\frac{\partial}{\partial p}\left[p^3\! \left( {
        \nu_S^{ab}} f_a + \frac{1}{2} {\nu_\parallel^{ab} } p
      \frac{\partial f_a}{\partial p} \right)\!\right],
      \label{eq:collop}
\end{equation}
where $\mathcal{L}$ represents scattering at constant energy~\cite{helander}, and $\nu_D^{ab}$, $\nu_S^{ab}$ and $\nu_\parallel^{ab}$ are the
deflection, slowing-down and parallel-diffusion frequencies which are well known in the limits of complete and no screening~\cite{helander}. In this Letter, we derive a generalization of these 
frequencies, taking into account the effect of reduced screening in the elastic
collisions between electrons and ions, as well as inelastic collisions
between fast and bound electrons.

{\em Elastic collisions.---}Elastic collisions with ions contribute to pitch-angle scattering (deflection frequency) through the
scattering cross section, which we evaluate in the Born
approximation. The Born approximation is valid for $\beta = v/c \gg
Z\alpha$ \cite{landauQM}, where $\alpha\approx 1/137$ is the
fine-structure constant. In the cross section, we neglect ion recoil for all ion species $j$ (since $m_e/m_j \ll 1$). Furthermore, since
we are interested in the effect on superthermal particles with $v\gg
v_{{\rm T}e} \gg v_{{\rm T}j}$, where $v_{{\rm T}a} = \sqrt{2 T_a/m_a}$ is the thermal
speed, we consider collisions with a narrow ion distribution:
$f_j(\mathbf{p}) = n_j \delta(\mathbf{p})$. The cross section then
takes the following form~\cite{landauQM,heitler}:
\begin{equation}
\frac{d\sigma_{ej}}{d\Omega} = \frac{r_0^2}{4 p^4}\! \lpa \frac{\cos^2(\theta/2) p^2 +1 }{\sin^4(\theta/2)}\rpa  \left|Z_j-F_j\left(q\right)\right|^2,
\label{eq:crossSection}
\end{equation}
where $r_0$ is the classical electron radius, $\theta$ is the deflection angle and $\mathbf{q}=2\textbf{p}\sin(\theta/2)/\alpha$.
The atomic form factor $F_j(q)$ is the Fourier transform of the electron
number density $\rho_{e,j}$ around an ion of species $j$: 
\begin{equation}
F_j(q) =  
\int \rho_{e,j}(r) e^{-i \mathbf{q\cdot r}/a_0}\,d\mathbf{r}\,,\label{eq:F(q)}
\end{equation}
where $a_0 = \hbar/(m_e c \alpha)$ is the Bohr radius. In the limit of
small $q$, the form factor approaches the number of bound
electrons $N_{e,j}=Z_j-Z_{0,j}$, giving a factor of $Z_{0,j}^2$ in the
cross section in Eq.~\eqref{eq:crossSection}, which corresponds to
complete screening. In the opposite limit of high momentum, the fast
oscillation of the integrand causes the form factor to vanish, giving a factor of
$Z^2_j$, representing no screening.

When deriving the generalized collision operator using Eqs.~\eqref{eq:crossSection} and \eqref{eq:F(q)}, we retain $x=
\sin{(\theta/2)}$ only to leading order since it can be shown that
small-angle collisions dominate. We do however allow $q = 2 x p /\alpha$
to be significant due to the large electron energies.  
Only the term including $\nu_D^{ab}$ in Eq.~\eqref{eq:collop} then persists:
\begin{align}
\nu_{D}^{ei} &= \nu_{D,\mathrm{\textsc{cs}}}^{ei} \Bigg( 1+
\frac{1}{Z_\mathrm{eff}}\sum_j \frac{n_j}{n_e}   
\frac{g_j(p)}{\ln\Lambda} \Bigg),
\label{eq:nuD}
\\ \label{eq:gFull}
g_j(p) &= \int_{0}^1 
x^{-1}\!\left\{[Z_j-F_j(q)]^2-Z_{0,j}^2 
\right\} dx,
\end{align}
where 
$\nu_{D,\mathrm{\textsc{cs}}}^{ei} = \tau_c^{-1}Z_\mathrm{eff} \gamma / p^3$ 
is the completely screened deflection frequency
for superthermal particles, 
and $\tau_c  = (4 \pi n_e c r_0^2 \ln\Lambda)^{-1}$ is the relativistic collision time. The effective charge is defined as
$Z_\mathrm{eff} = \sum_j n_j Z_{0,j}^2/n_e$, where $n_e$ represents the density of free electrons.  
The lower integration limit in $g_j(p)$ (which is formally $1/\Lambda \ll 1$) has been set to zero, since the integrand is finite.  We model the Coulomb logarithm according to $ \ln \Lambda^{ee} = \ln\Lambda_0 +
\frac{1}{k}\ln\left\{1+[2(\gamma\!-\!1)/p^2_{{\rm T}e}]^{k/2} \right\}$ and $ \ln \Lambda^{ei} = \ln\Lambda_0 +
\frac{1}{k}\ln\left[1+(2 p/p_{{\rm T}e})^k \right]$, where
$p_{{\rm T}e}$ is the thermal momentum. The parameter $k=5$ is chosen to give a smooth transition between the 
low-energy formula
$\ln\Lambda_0 = 14.9-0.5 \ln(n_e[10^{20}\,{\mathrm m}^{-3}])+\ln
(T_e[{\mathrm{keV}}])$~\cite{wesson} and the high-energy formula from Refs.~\cite{Gould,SolodovBetti}.

To calculate the form factor, the electron charge density of the ion
can be obtained via e.g.\! DFT calculations; in this work we have
used the numerical tool {\sc exciting}~\cite{excitingrevp}. 
Our calculations show that the form factor can be well described by a
single-parameter model of the same form as that obtained from the
Thomas--Fermi model by Kirillov et al.~\cite{Kirillov}:
$ F_{j,\mathrm{\textsc{tf-dft}}}(q) = N_{e,j}/[1+(q a_j)^{3/2}]$.  
This model, which we denote the Thomas-Fermi--DFT (TF-DFT) model, gives 
\begin{equation}
g_j(p) = \frac{2}{3}(Z_j^2-Z_{0,j}^2)\ln(y_j^{3/2}+1)-\frac{2}{3} \frac{N_{e,j}^2 y_j^{3/2}}{y_j^{3/2}+1},
\label{eq:g}
\end{equation}
where $y_j = 2 a_j p/\alpha$. Note that the limit of complete screening ($g_j(p)\rightarrow 0$) is reached  as  $p\rightarrow 0$ or for  $Z_j\!=\!Z_{0,j}$.  The parameter $a_j$ -- the effective ion size in units of Bohr radii -- depends on the ion species and ionization degree, and was determined by fitting
$g_j$ in Eq.~\eqref{eq:g} to Eq.~\eqref{eq:gFull} evaluated with the DFT output. For example, we obtain 
 $a_{\mathrm{Ar}} \!=\! 0.353$,
 $a_{\mathrm{Ar}^+} \!=\! 0.329$, 
 $a_{\mathrm{Ar}^{2+}}\!=\! 0.306$, 
 $a_{\mathrm{Ar}^{3+}} \!=\!0.283$, 
 $a_{\mathrm{Ar}^{4+}} \!=\! 0.260$, 
 $a_{\mathrm{Ar}^{5+}} \!=\! 0.238$, 
 $a_{\mathrm{Xe}^+} \!=\! 0.238$, 
and
 $a_{\mathrm{Be}^{+}}\!=\!0.414$. 

\begin{figure}\centering
\includegraphics[width=0.45\textwidth,trim={0.4cm 0.7cm 0.5cm 0.3cm}]{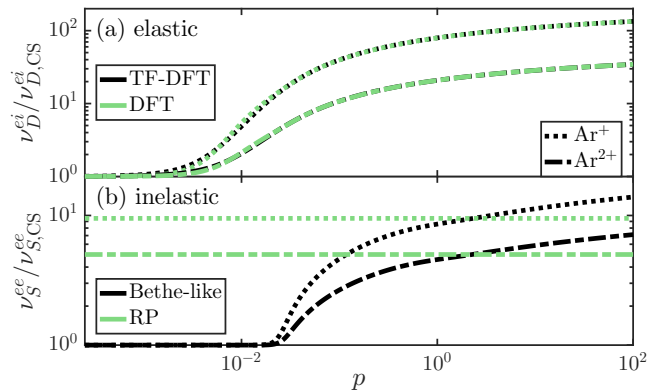}
\caption{ (a) The deflection frequency and (b) the slowing-down
  frequency as a function of the incoming-electron 
  momentum,
  normalized to the completely screened collision frequencies.  The models
  employed here (TF-DFT and Bethe-like) are plotted in black, while
  the full DFT model and the approximate RP model are shown in green.
Note in (a) the lines overlay almost exactly.  A pure argon plasma 
(of either Ar$^{+}$ (dotted) or Ar$^{2+}$ (dash-dotted)),  
  with $T\!=\!10\,$eV and $n_\mathrm{Ar}\!=\!10^{20}\,$m$^{-3}$  was
  assumed, giving $\ln\Lambda_0^{+}\!=\!9.9$ and $\ln\Lambda_0^{2+}\!=\!10.3$.}
    \label{fig:nu} 
\end{figure}

The TF-DFT model agrees well with the prediction of full
DFT simulations.  Figure~\ref{fig:nu}~(a) shows the
energy-dependent enhancement of the deflection frequency normalized to the  completely screened value,
together with the fit given in Eq.~\eqref{eq:g} for singly and doubly-ionized argon. The deflection
frequency is already almost two orders of magnitude higher than the
corresponding complete-screening value at electron energies of a
few hundred keV ($p \approx 1$). 

{\em Inelastic collisions.---}The energy loss in electron-electron collisions is described by the Bethe stopping-power formula~\cite{bethe,jackson}, which modifies the slowing-down frequency $\nu_S^{ee}$ describing collisional drag according to $ \nu_S^{ee} = \nu_{S,\mathrm{\textsc{cs}}}^{ee} 
\left[ 1+
\sum_j (n_j N_{e,j}/n_e \ln\Lambda)
\left( \ln h_j-\beta^2\right)\right],$
where
$\nu_{S,\mathrm{\textsc{cs}}}^{ee} = \tau_c^{-1} \gamma^2/p^3$
is the completely screened, superthermal slowing-down frequency,    $h_j = p\sqrt{\gamma-1}/I_j$,  and $I_j$ is the mean excitation energy of the ion, normalized to the electron rest energy.
In this work, the numerical values of $I_j$ for different ion species were obtained from Ref.~\cite{sauer2015}.
This model is valid for $\gamma-1 \gg I_j$, corresponding to $p \gtrsim 0.03$ for both singly and doubly ionized argon.
We provide here an interpolation formula, from matching the above to the low-energy
asymptote corresponding to complete screening, which we will refer to as
the \emph{Bethe-like model}:
\begin{align}
\nu_S^{ee}  = & \nu_{S,\mathrm{\textsc{cs}}}^{ee}
\bigg\{ 1+ 
 \!\sum_j\! \frac{n_j N_{e,j}}{n_e \ln \Lambda} 
\left[\frac{1}{k}
\ln\!\left( 1+ \!h_j^k\right)
-\beta^2\right] \!\!\bigg\}.
\label{eq:nuS}
\end{align}
As in our model of $\ln\Lambda$, we set $k=5$.
 
Figure~\ref{fig:nu}~(b) shows the enhancement of the slowing-down
frequency as a function of the electron energy. 
Note
that already around a few tens of keV the enhancement using the
Bethe-like model (black) is an order of magnitude. The transition
between the Bethe equation and the low-energy limit can be clearly
seen at $p \approx 0.02$.
It is instructive to compare
these results to the \emph{Rosenbluth--Putvinski} (RP) rule of
thumb that the effect of inelastic collisions can be modeled by adding half of the bound electrons to the free electron
density~\cite{rosenbluthPutvinski}:
$
\nu_{S,\mathrm{\textsc{rp}}}^{ee} \approx \nu_{S,\mathrm{\textsc{cs}}}^{ee}
\big[
 1 + 
\frac{1}{2}\sum_jn_jN_{e,j}/n_{e}
\big]$. 
This approximation 
(green line in Fig.~\ref{fig:nu}~(b)) leads to a much greater 
enhancement than the full formula up to $p\simeq 0.1$. This region in momentum space is important, since runaway generation is sensitive to the dynamics at the critical momentum $p_c$, which often is in the region $p_c \lesssim 0.1$.
The effect of inelastic collisions on the electron-electron deflection
frequency $\nu_D^{ee}$ does not follow from the stopping-power calculation,
but is of order $X^{-1}$ smaller than $\nu_D^{ei}$ and can be ignored for low ionization degrees.

{\em Numerical simulations.---}The generalized collision operator
presented here, consisting of Eqs.~\eqref{eq:collop}, \eqref{eq:nuD},
\eqref{eq:g}, and \eqref{eq:nuS}, has been implemented in the
numerical tool \textsc{code} \cite{CODEpaper2014,Stahl2015,Stahl2016}, which we use to solve the spatially-homogeneous kinetic 
equation for electrons in 2D momentum space,
including electric-field acceleration, collisions and synchrotron-radiation reaction losses.

We demonstrate the effects of reduced screening by investigating the decay phase of the runaway evolution. In the scenario considered, an electron distribution with an energetic runaway tail (with average energy  $7.8\,$MeV), produced by a strong electric field, was used as the initial state. 
During the simulation of the decay phase, the weak electric field $E = 2 E_c$ was used, which is well below the {\em effective} critical field if reduced screening effects are taken into account. To isolate the effect of reduced screening, avalanche runaway  generation (which is unimportant on the simulation timescale~\cite{rosenbluthPutvinski}) was neglected, and bremsstrahlung  (which can sometimes be an important energy-loss mechanism~\cite{BakhtiariPRL2005,EmbreusBrems2016}) does not affect the dynamics significantly at the energies considered.

\begin{figure}
\centering
\includegraphics[width=0.46\textwidth,trim={0.4cm 0.7cm 0.2cm 0.3cm}]{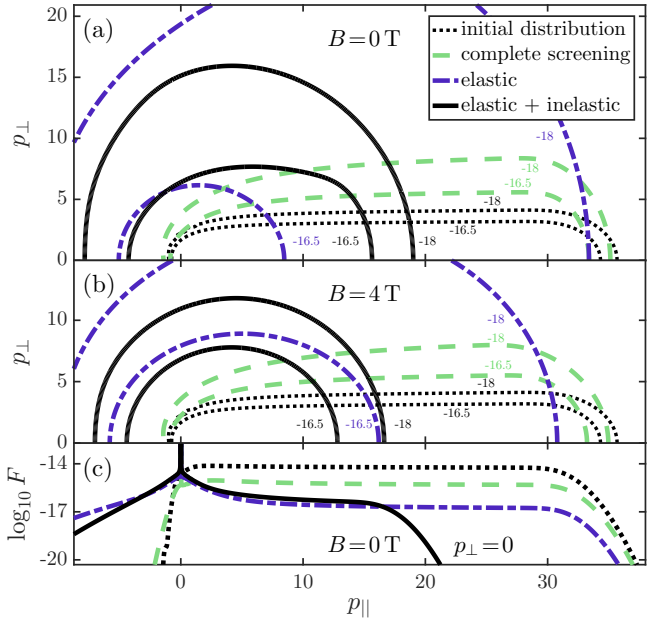}
\caption{Contours of the distribution after $25\,$ms of collisional deceleration from an initial beam-like state (dotted black), with and without screening effects. The limit of complete screening for both $\nu_D^{ei}$ and $\nu_S^{ee}$ (dashed green) is shown, together with the TF-DFT model for $\nu_D^{ei}$ but unmodified $\nu_S^{ee}$ (dash-dotted blue), and with both the TF-DFT and Bethe-like models (solid black). The
  contours $\log_{10}(F) =-16.5$ and $-18$ are shown, where $F\! =\! (2 \pi m_e T)^{3/2} f_e/n_e$, with (a)~$B\!=\!0\,$T and (b)~$B\!=\!4\,$T. A parallel cut through the distribution of (a) is shown in (c).  Parameters: $T\!=\!10\,$eV,
  $n_\mathrm{H}\!=\!10^{20}$ m$^{-3}$ and Ar$^{+}$ with density
  $n_\mathrm{Ar}\! =\! n_\mathrm{H}$, $E\! =\! 2E_c$.}
    \label{fig:contours}
\end{figure}

The effect of reduced screening on the electron distribution is
shown in Fig.~\ref{fig:contours}, (a) without and (b) with
synchrotron radiation, in a plasma with equal amounts of singly ionized argon and hydrogen.  
As shown in the figure,
the enhanced deflection frequency (dash-dotted blue) acts
to make the distribution function isotropic, as can be seen by a comparison to the completely screened case (dashed green).  
Energy losses are induced when, in addition, the slowing-down model of inelastic collisions with bound electrons is included (solid black).
The synergy with synchrotron radiation further enhances the energy loss as shown in Fig.~\ref{fig:contours}(b). 
The one-dimensional plot in Fig.~\ref{fig:contours}(c) offers a complementary view of the  distribution function; note that the maximum runaway energy is lower if inelastic collisions are included, compared to only considering elastic collisions.

\begin{figure}
\centering
\includegraphics[width=0.46\textwidth,trim={0.2cm 0.7cm 0.2cm 0.3cm}]{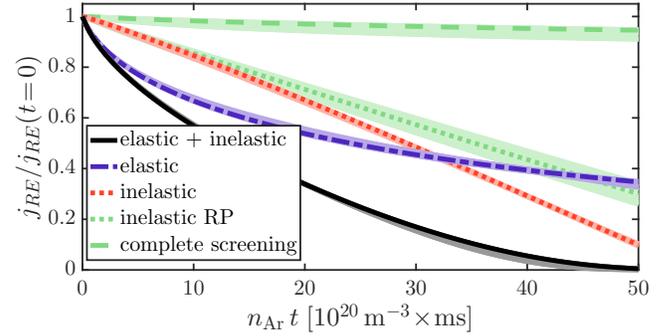}
\caption{Decay of the runaway-electron current as a function of argon density and time. The full model, with both the TF-DFT and
  Bethe-like contributions (solid black), as well as the $\nu_D^{ei}$ (dash-dotted blue) and $\nu_S^{ee}$ (dotted red) models separately, are shown.  
These are compared to the RP model for inelastic collisions (dotted green) and the limit of complete screening (dashed green).
 The initial distribution of  Fig.~\ref{fig:contours} was used, assuming an H plasma with $n_\mathrm{H}=10^{20}\,$m$^{-3}$, and Ar$^{+}$ impurities with $n_\mathrm{Ar}\!=\!10\:n_\mathrm{H}$; the light bands show the range of results obtained by varying the argon density such that 
  $n_\mathrm{Ar}\!\in\! [0.5\:n_\mathrm{H},100\: n_\mathrm{H}]$.  $T\!=\!10$~eV, $E\!=\!2E_c$ and $B\!=\!2\,$T.}
    \label{fig:jRE}
  \end{figure}

The changes to the distribution function presented in
Fig.~\ref{fig:contours} have significant implications for the decay of
a runaway-electron current, as shown in Fig.~\ref{fig:jRE}. Since the
decay time to a good approximation is proportional to
$1/n_\mathrm{Ar}$ for $n_\mathrm{Ar}\!\gtrsim\! n_\mathrm{H}$
(i.e.~when $\nu_{D}^{ei} \!\gg\! \nu_{D,\mathrm{\textsc{cs}}}^{ei}$
and $\nu_{S}^{ee} \!\gg\! \nu_{S,\mathrm{\textsc{cs}}}^{ee}$), the
runaway-electron current is shown as a function of
$n_{\mathrm{Ar}}\,t$. 
With the full model of reduced screening, the
current-decay time is reduced by orders of magnitude compared to the
complete-screening model.  The effect on the current is due to the
combination of the inelastic and the elastic collisions: elastic
collisions are most important when the distribution is narrow in
pitch angle, while inelastic collisions will dominate at later times,
when the distribution is sufficiently isotropic so that elastic
scattering is less efficient. The RP model (which was applied for momenta $p\!>\!10\:p_{{\rm T}e}$) underestimates the decay rate resulting from inelastic collisions, and shows a widely different current evolution compared to the full model.

  The bands in Fig.~\ref{fig:jRE} represent an impurity
    density scan over two orders of magnitude, showing only very
    slight variation from the linear relationship between decay rate
    and $n_{\mathrm{Ar}}$.  Similarly, the current decay is
    insensitive to the electric field as long as it is significantly
    lower than the effective critical field. Although the relative
    importance of elastic and inelastic collisions is influenced by
    the width of the distribution function in pitch-angle, the overall
    effect on the current decay is not strongly affected as long as
    the initial distribution is forward-beamed. 
        
While the current decay shown in Fig.~\ref{fig:jRE} is a robust result when the  electric field is constant, it neglects the inductive coupling between the current, $I$, and the electric field in a tokamak, which is allowed when $I\!\lesssim\!250\,$kA~\cite{breizman2014}. 
The opposite, highly inductive limit leads to a current decay rate proportional to the critical electric field~\cite{breizman2014}:
$dI/dt\!=\!2 \pi R E_{c}^{\rm eff}/L,$
where $L\!\sim\!\mu_0 R$ is the self-inductance and $R$ is the major
radius. To
calculate $E_{c}^{\rm eff}$, which is 
increased due to reduced screening compared to the classical value
$E_c$, we assume fast pitch-angle dynamics following the procedure in
Refs.~\cite{lehtinen,AleynikovPRL}.  For an H plasma with up to triply ionized Ar impurities and densities in the range $n_\mathrm{Ar}\!\gtrsim\! 0.1 n_\mathrm{H}$, and neglecting synchrotron losses ($B [{\rm T}]^2\! \lesssim \! n_{\rm Ar} [10^{18} \,{\rm m}^{-3}]$), we obtain
\begin{equation}
    \frac{E_{c}^{\rm eff}}{E_c}\approx 1+\frac{1}{\ln\Lambda_0}\left(7-\ln \sqrt{T_{\rm eV}}+240 \frac{n_{\rm Ar,tot}}{n_e} \right). 
\end{equation}
This large enhancement of the critical electric field has significant contributions from elastic and inelastic collisions, implying that both effects are important for the runaway dynamics regardless of the inductance.
    
     {\em Conclusions.---} In this Letter, we give convenient analytical
  expressions for the effect of reduced screening on the collisional
  deflection and slowing-down frequencies, derived from first
  principles.  The model is formally correct where the Born
  approximation is valid ($v/c \gg Z\alpha$), but is applicable for
  all electron energies due to a matching to the completely-screened
  low-energy limit.  For the first time, we investigate the electron
  dynamics using kinetic simulations, and find that the
  reduced-screening effect of bound electrons has a large impact on
  the distribution function of runaway electrons. The enhancement of
  both collisional drag and pitch-angle scattering lead to significant
  energy loss, the latter due to the increased synchrotron radiation. 
We provide a formula for the effective critical field --  the threshold for runaway generation -- which can be used to predict the runaway-current decay time in tokamaks.
   Our results indicate that runaway beams will be strongly damped even
  in the presence of weakly ionized impurities, in agreement with
  experiments~\cite{Hollmann2015}.  
  Given the impact of collisions with screened nuclei on the dynamics
  of runaway electrons, the effects detailed here should be considered
  in all situations where fast electrons interact with partially
  ionized impurities, e.g.\! lightning discharges, tokamak
    disruptions and laser-plasma interaction.

\begin{acknowledgments}
This work was supported by the Swedish Research Council
(Dnr.~2014-5510), the Knut and Alice Wallenberg Foundation and the
European Research Council (ERC-2014-CoG grant 647121). The authors are
grateful for stimulating discussions with Matt Landreman, George Wilkie and Christian
Forss\'en.
\end{acknowledgments}

\bibliography{references} 

\end{document}